\documentclass[12pt]{article}
\usepackage[round]{natbib}
\usepackage{graphicx,url,amsmath,amssymb,sectsty,mathbbol,rotating,url}
\allsectionsfont{\centering}
\renewcommand\footnotemark{}

\title{\textbf{Stability Indicators in Network Reconstruction}}
\author{
GIUSEPPE JURMAN$^{1}$, MICHELE FILOSI$^{1,2}$, ROBERTO VISINTAINER$^{1,3}$, \\
SAMANTHA RICCADONNA$^{1}$ and CESARE FURLANELLO$^{1}$
\footnote{$^{1}$ Fondazione Bruno Kessler (FBK), Trento, Italy\newline\indent\indent$^{2}$ Centre for Integrative Biology (CIBIO), University of Trento, Trento, Italy\newline\indent\indent$^{3}$ Department of Information Engineering and Computer Science (DISI), University of Trento, Trento } }
\date{}

\begin{document}
\maketitle

\section*{ABSTRACT}
\textbf{ 
The number of algorithms available to reconstruct a biological network from a dataset of high-throughput measurements is nowadays overwhelming, but evaluating their performance when the gold standard is unknown is a difficult task.
Here we propose to use a few reconstruction stability tools as a quantitative solution to this problem.
We introduce four indicators to quantitatively assess the stability of a reconstructed network in terms of variability with respect to data subsampling.
In particular, we give a measure of the mutual distances among the set of networks generated by a collection of data subsets (and from the network generated on the whole dataset) and we rank nodes and edges according to their decreasing variability within the same set of networks.
As a key ingredient, we employ a global/local network distance combined with a bootstrap procedure.
We demonstrate the use of the indicators in a controlled situation on a toy dataset, and we show their application on a miRNA microarray dataset with paired tumoral and non-tumoral tissues extracted from a cohort of 241 hepatocellular carcinoma patients.
}
\vskip.5cm
\textbf{Key words:} Network inference, Network comparison, High-throughput data

\section{INTRODUCTION}
\label{sec:intro}
The problem of inferring a biological network structure starting from a set of high-throughput measurements (\textit{e.g.} gene expression arrays) has been positively answered by a huge number of deeply different solutions published in literature in the last fifteen years.
Nonetheless, network reconstruction suffers from being a underdetermined problem, being the number of interactions highly larger than the number of independent measurements \citep{desmet10advantages}: thus any algorithm has to look for a compromise between accuracy and feasibility, allowing simplifications that inevitably mine the precision of the final outcome, for instance including a relevant number of false positive links \citep{kamburov12intscore}.
This makes the inference problem "a daunting task" \citep{baralla09inferring}, not only in terms of devising an effective algorithm, but also in terms of quantitatively interpreting the obtained results. 
In general, the reconstruction accuracy is far from being optimal in many situations with the presence of several pitfalls \citep{meyer11verification}, related to both the methods and the data \citep{he09reverse}, with the extreme situation of many link prediction being statistically equivalent to random guesses \citep{prill10towards}. 
In particular, the size (and the quality) of the available data play a critical role in the inference process, as widely acknowledged \citep{logsdon10gene,gillis11role,miller12identifying}.
All these considerations support deeming network reconstruction a still unsolved problem \citep{szederkenyi11inference}.

Despite the ever rising number of available algorithms, only recently efforts have been carried out towards an objective comparison of network inference methods also highlighting current limitations \citep{altay10revealing,krishnan07indeterminacy} and relative strengths and disadvantages \citep{madhamshettiwar12gene}.
Among those, it is worthwhile mentioning the international DREAM challenge \citep{marbach10revealing}, whose key result in the last edition advocated integration of predictions from multiple inference methods as an effective strategy to enhance performances taking advantage from the different algorithms' complementarity \citep{desmet10advantages}.
Nevertheless, the algorithm uncertainty has been so far assessed only in terms of performance, i.e. distance of the reconstructing network from the ground truth, wherever available, while not much has been instead investigated with respect to the stability of the methods. 
This can be of particular interest when no gold standard is available for the given problem, and thus there is no chance to evaluate the algorithm's accuracy, leaving the stability as the sole rule of thumb for judging the reliability of the obtained network.
Here we propose to tackle the issue by quantifying inference variability with respect to data perturbation, and, in particular, data subsampling.
If a portion of data is randomly removed before inferring the network, the resulting graph is likely to be different from the one reconstructed from the whole dataset and, in general, different subsets of data would generate different networks.
Thus, in the spirit of applying reproducibility principles to this field, one has to accept the compromise that the inferred/non inferred links are just an estimation, lying within a reasonable probability interval. 
In brief, we aim at proposing a set of four indicators allowing the researcher to quantitatively evaluate the reliability of the inferred/non-inferred links.
In detail, we quantitatively assess, for a given ratio of removed data and for a give number of resampling, the mutual distances among all inferred networks and their distances to the network generated by the whole dataset, with the idea that, the smaller the average distance, the stabler the network.
Moreover, we provide a ranked list of the stablest links and nodes, where the rank is induced by the variability of the link weight and the node degree across the generated networks, the less variable being the top ranked.
As a network distance we employ the HIM distance \citep{jurman12glocal}, which represents a good compromise between local (link-based) and global (structure-based) measure of network comparison.

As a first testbed in a controlled situation the four indicators are computed on a synthetic dataset for different instances of a correlation network with different measures, highlighting the impact of a FDR filter on the network reconstruction method.
Finally, we show the use of the stability measures in comparing the relevance networks inferred on a miRNA microarray dataset with paired tissues extracted from a cohort of 241 hepatocellular carcinoma patients \citep{budhu08identification}. 
Data have two phenotypes, related to disease (tumoral or non-tumoral tissues) and to patient's sex (male or female), allowing the construction of four different networks, displaying different levels of stability.

Due to the relevant computational workload, all the analysis were run as R and Python scripts on multicore workstations and on the FBK HPC facility Kore Linux cluster. 

\section{METHODS}
\label{sec:methods}
Before defining the four stability indicators we briefly summarize the main definitions and properties of the HIM network distance.

\subsection{HIM Network Distance}
\label{ssec:him}
The HIM distance \citep{jurman12glocal} is a metric for network comparison combining an edit distance (Hamming \citep{tun06metabolic,dougherty10validation}) and a spectral one (Ipsen-Mikhailov \citep{ipsen02evolutionary}).
As discussed in \citep{jurman10introduction}, edit distances are local, that is they focus only on the portions of the network interested by the differences in the presence/absence of matching links.  
Spectral distances evaluate instead the global structure of the compared topologies, but they distinguish isomorphic or isospectral graphs, which can correspond to quite different conditions within the biological context.
Their combination into the HIM distance represents an effective solution to the quantitative evaluation of network differences.

Let $\mathcal{N}_1$ and $\mathcal{N}_2$ be two simple networks on $N$ nodes, described by the corresponding adjacency matrices $A_1$ and $A_2$, with $a^{(1)}_{ij}, a^{(2)}_{ij}\in\mathcal{F}$, where $\mathcal{F}=\mathbb{F}_2=\{0,1\}$ for unweighted graphs and $\mathcal{F}=[0,1]$ for weighted networks. 
Denote then by $\mathbb{I}_N$ the identity $N\times N$ matrix $\mathbb{I}_N = \left( \begin{smallmatrix} 1&0&\cdots & 0 \\ 0&1&\cdots&0 \\ &\cdots \\ 0&0&\cdots &1  \end{smallmatrix} \right)$, by $\mathbb{1}_N$ the unitary $N\times N$ matrix with all entries equal to one and by $\mathbb{0}_N$ the null $N\times N$ matrix with all entries equal to zero. 
Finally, denote by $\mathcal{E}_N$ the empty network with $N$ nodes and no links (with adjacency matrix $\mathbb{0}_N$) and by $\mathcal{F}_N$ the undirected full network with $N$ nodes and all possible $N(N-1)$ links (whose adjacency matrix is $\mathbb{1}_N-\mathbb{I}_N$).

The definition of the Hamming distance is the following:
\begin{displaymath}
\textrm{Hamming}(\mathcal{N}_1,\mathcal{N}_2) = \sum_{1\leq i\not = j\leq N} \vert A^{(1)}_{ij} - A^{(2)}_{ij} \vert\ .
\end{displaymath}
To guarantee independence from the network dimension (number of nodes), we normalize the above function by the factor $\overline{\eta}=\textrm{Hamming}(\mathcal{E}_N,\mathcal{F}_N)=N(N-1)$:
\begin{equation}
\label{eq:hamming}
H(\mathcal{N}_1,\mathcal{N}_2) = \frac{1}{N(N-1)} \sum_{1\leq i\not = j\leq N} \vert A^{(1)}_{ij} - A^{(2)}_{ij} \vert\ .
\end{equation}
When $\mathcal{N}_1$ and $\mathcal{N}_2$ are unweighted networks, $H(\mathcal{N}_1,\mathcal{N}_2)$ is just the fraction of different matching links (over the total number $N(N-1)$ of possible links) between the two graphs.
In all cases, $H(\mathcal{N}_1,\mathcal{N}_2)\in [0,1]$, where the lower bound $0$ is attained only for identical networks $A_1=A_2$ and the upper bound $1$ is reached whenever the two networks are complementary $A_1+A_2=\mathbb{1}_N-\mathbb{I}_N=\left( \begin{smallmatrix} 0&1&\cdots & 1 \\ 1&0&\cdots&1 \\ &\cdots \\ 1&1&\cdots &0  \end{smallmatrix} \right)$.

Among spectral distances, we consider the Ipsen-Mikhailov distance IM which has been proven to be the most robust in a wide range of situations \citep{jurman10introduction}. 
Originally introduced in \citep{ipsen02evolutionary} as a tool for network reconstruction from its Laplacian spectrum, the definition of the Ipsen-Mikhailov metric follows the dynamical interpretation of a $\textrm{N}$--nodes network as a $\textrm{N}$--atoms molecule connected by identical elastic strings, where the pattern of connections is defined by the adjacency matrix of the corresponding network.  
The dynamical system is described by the set of $N$ differential equations 
\begin{equation}
\label{eq:ipsen_model}
\ddot{x}_i+\sum_{j=1}^N A_{ij}(x_i-x_j)=0\quad\textrm{for\ }i=0,\cdots,N-1\ .
\end{equation}
We recall that the Laplacian matrix $L$ of an undirected network is defined as the difference between the degree $D$ and the adjacency $A$ matrices $L=D-A$, where $D$ is the diagonal matrix with vertex degrees as entries.
$L$ is positive semidefinite and singular \citep{chung97spectral,atay06network,spielman09spectral,tonjes09perturbation,atay06network}, so its eigenvalues are $0 = \lambda_0 \leq \lambda_1\leq \cdots\leq \lambda_{N-1}$.
The vibrational frequencies $\omega_i$ for the network model in Eq.~\ref{eq:ipsen_model} are given by the eigenvalues of the Laplacian matrix of the network: $\lambda_i = \omega^2_i$, with $\lambda_0=\omega_0=0$. 
The spectral density for a graph as the sum of Lorentz distributions is defined as 
\begin{displaymath}
\rho(\omega,\gamma)=K\sum_{i=1}^{N-1} \frac{\gamma}{(\omega-\omega_i)^2+\gamma^2}\ ,
\end{displaymath}
where $\gamma$ is the common width and $K$ is the normalization constant defined as
\begin{displaymath}
K = \frac{1}{\gamma\displaystyle{\sum_{i=1}^{N-1} \int_0^\infty \frac{\textrm{d}\omega}{(\omega-\omega_i)^2+\gamma^2} }}\ ,
\end{displaymath}
so that $\displaystyle{\int_0^\infty \rho(\omega,\gamma)\textrm{d}\omega =1}$.
The scale parameter $\gamma$ specifies the half-width at half-maximum, which is equal to half the interquartile range. 
Then the spectral distance $\epsilon_\gamma$ between two graphs $G$ and $H$ on $N$ nodes with densities $\rho_G(\omega,\gamma)$ and $\rho_H(\omega,\gamma)$ can then be defined as 
\begin{displaymath}
\epsilon_\gamma(G,H) = \sqrt{\int_0^\infty \left[\rho_G(\omega,\gamma)-\rho_H(\omega,\gamma)\right]^2 \textrm{d}\omega}\ .
\end{displaymath}
The highest value of $\epsilon_\gamma$ is reached, for each $N$, when evaluating the distance between $\mathcal{E}_N$ and $\mathcal{F}_N$.
Defining $\overline{\gamma}$ as the (unique) solution of 
\begin{displaymath}
\epsilon_\gamma(\mathcal{E}_N, \mathcal{F}_N) = 1\ , 
\end{displaymath}
we can now define the normalized Ipsen-Mikahilov distance as
\begin{displaymath}
\textrm{IM}(G,H)=\epsilon_{\overline\gamma}(G,H) = \sqrt{\int_0^\infty \left[\rho_G(\omega,\overline{\gamma})-\rho_H(\omega,\overline{\gamma})\right]^2 \textrm{d}\omega}\ ,
\end{displaymath}
so that $\textrm{IM}(G,H)\in [0,1]$ with upper bound attained only for $(G,H)=(\mathcal{E}_N,\mathcal{F}_N)$.
\begin{figure}[!t]%
\begin{center}
\begin{tabular}{cc}
\begin{minipage}[!t]{0.22\textwidth}
\includegraphics[width=1\textwidth]{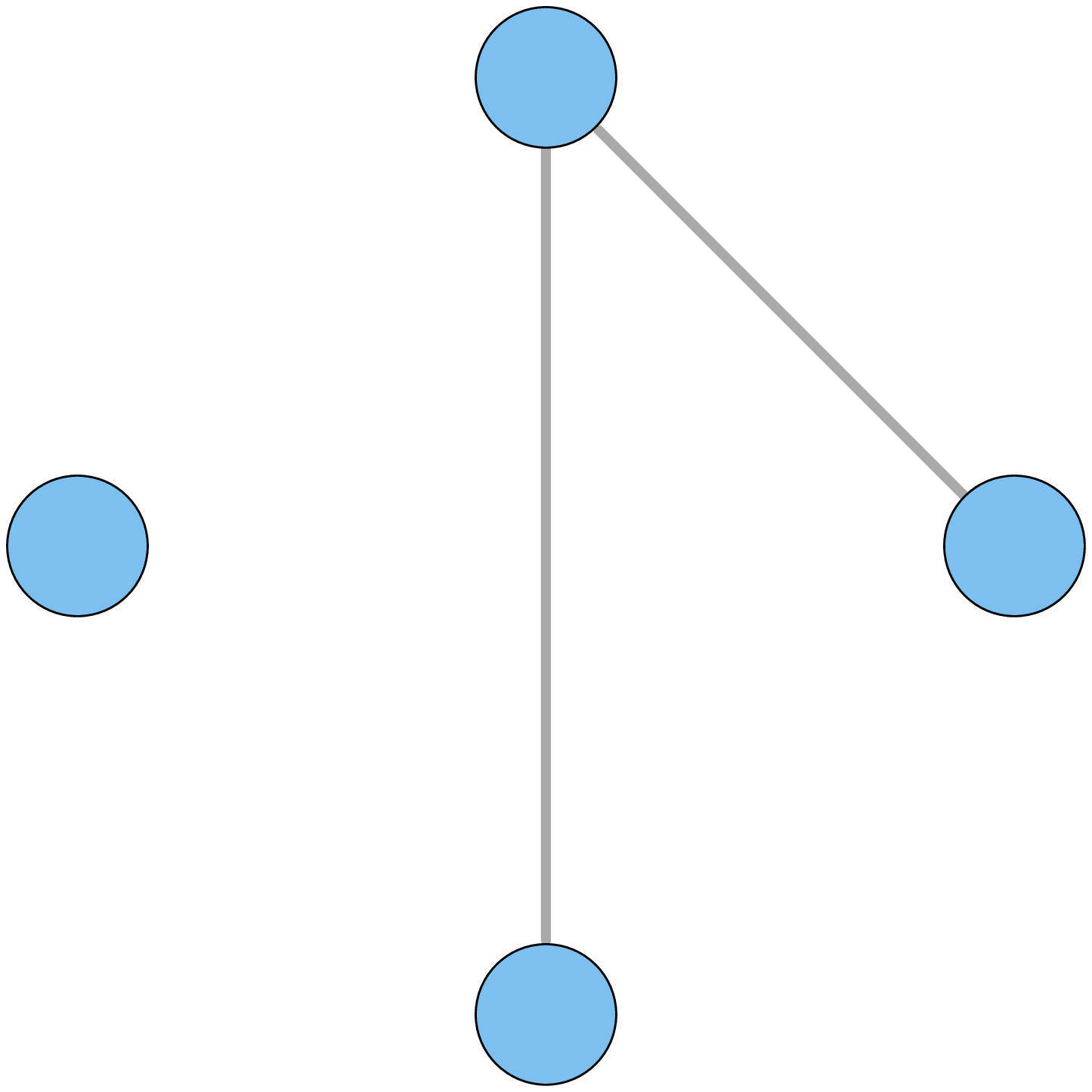}\\
\vspace{-0.5cm}\includegraphics[width=1\textwidth]{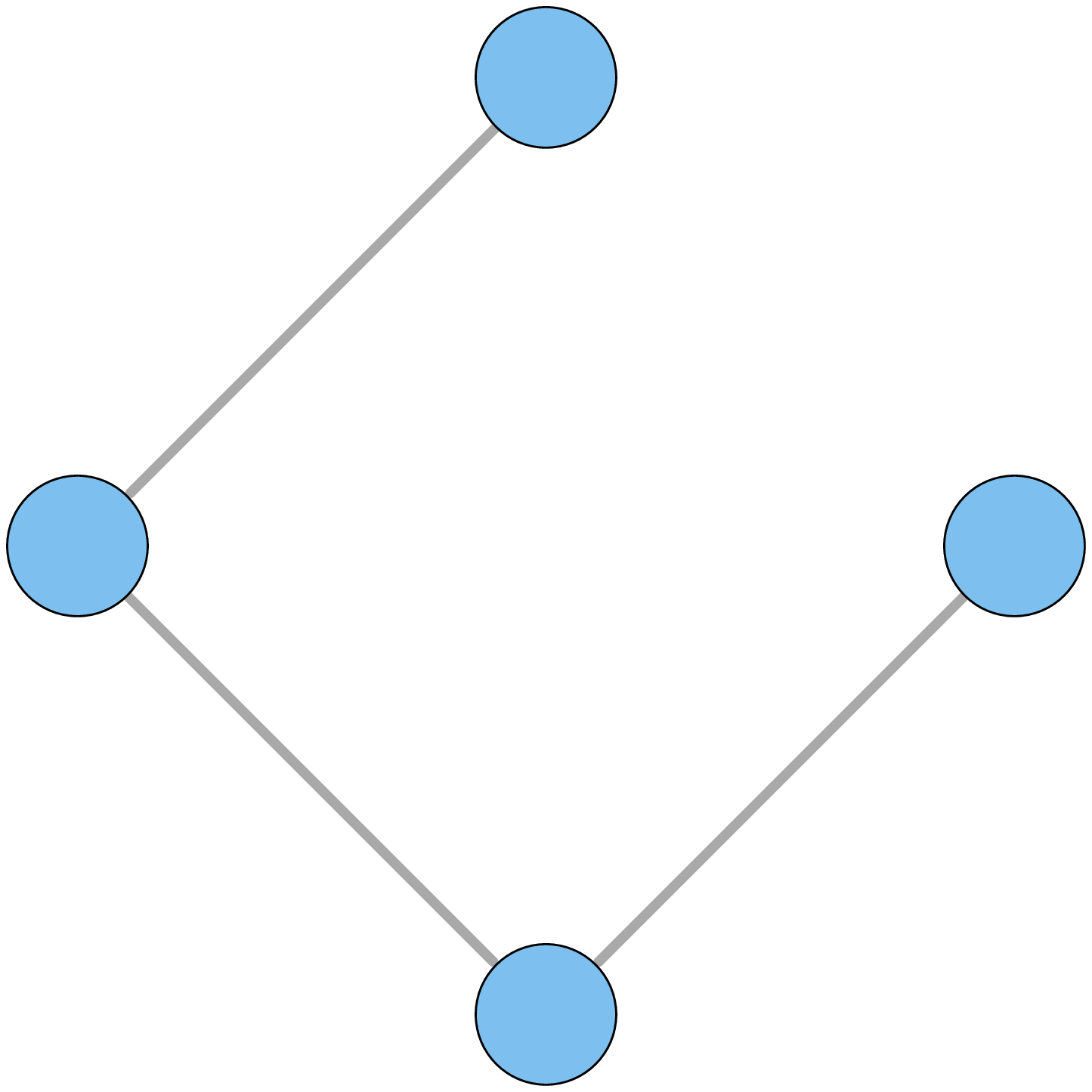}\\
\end{minipage}
&
\begin{minipage}[!t]{0.56\textwidth}
\includegraphics[width=1\textwidth]{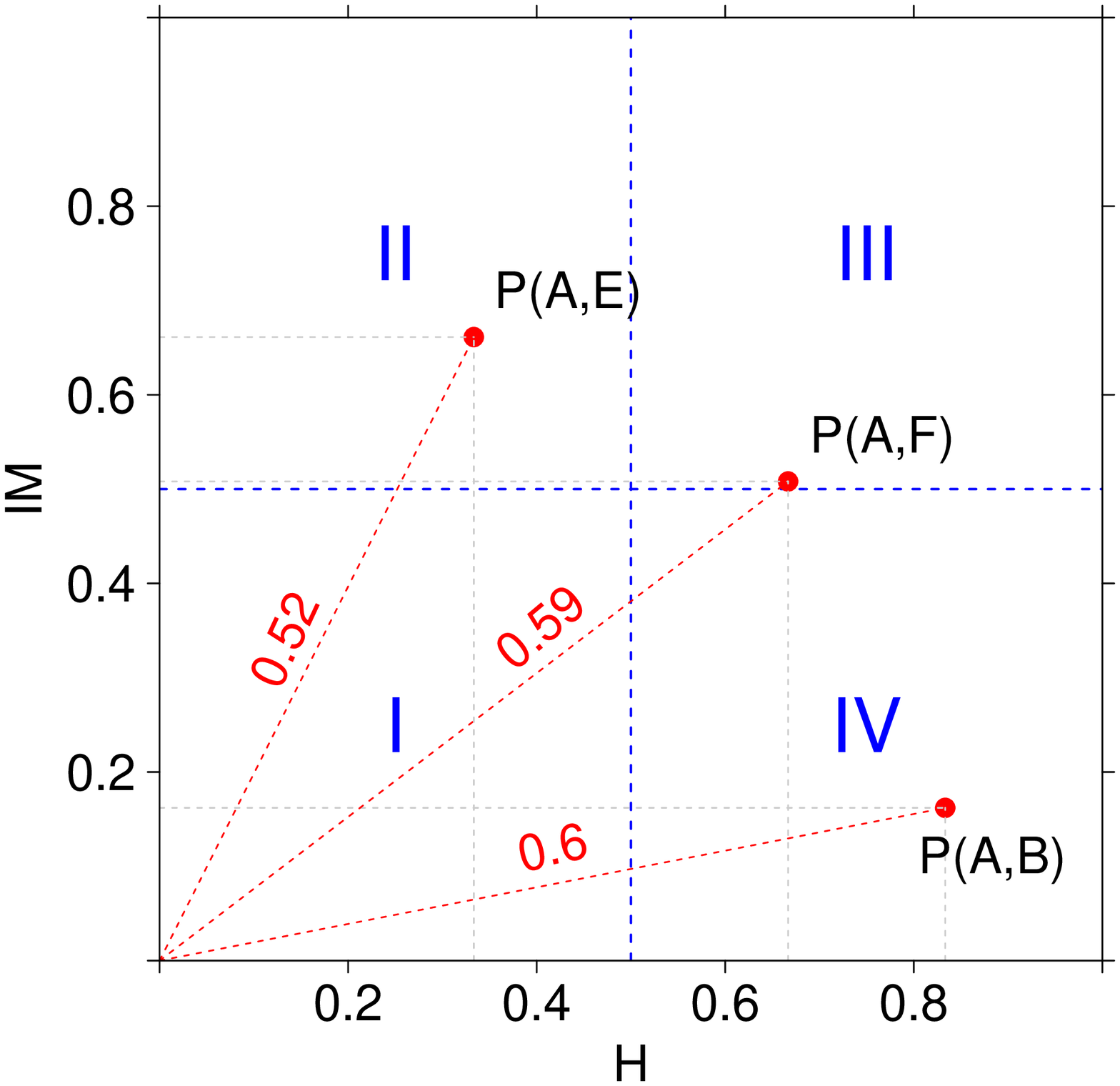}
\end{minipage}\\
(a)&(b)\\
\end{tabular}
\end{center}
\caption{An example of HIM distance. (a) Network A (top) and Network B (bottom); (b) Representation of the HIM distance in the Ipsen-Mikhailov and Hamming distance space between networks A versus B, C and D, where C is the fully connected network and D is the empty one.  }
\label{fig:him}
\end{figure}
Finally, the HIM distance is defined as the product metric of the normalized Hamming distance H and the normalized Ipsen-Mikhailov IM distance, normalized by the factor $\sqrt{2}$ to set its upper bound to 1:
\begin{displaymath}
\textrm{HIM}(N_1,N_2)=\frac{1}{\sqrt{2}}\sqrt{\text{H}(N_1,N_2)^2+\text{IM}(N_1,N_2)^2}
\end{displaymath}
We can represent the HIM distance in the $[0,1] \times [0,1]$ Hamming/Ipsen-Mikhailov space, where a point $P(x,y)$ represents the distance between two networks $N_1$ and $N_2$ whose coordinates are $x=\text{H}(N_1,N_2)$ and $y=\text{IM}(N_1,N_2)$ and the norm of $P$ is $\sqrt{2}$ times the HIM distance $\text{HIM}(N_1,N_2)$. 
The same holds for weighted networks, provided that the weights range in $[0;1]$.
In Fig.~\ref{fig:him} we provide an example of this representation of the HIM distance between networks of four nodes.
Roughly splitting the Hamming/Ipsen-Mikhailov space into four main zones I,II,III,IV as in Figure \ref{fig:him}, we can say that two networks whose distances correspond to a point in zone I are quite close both in terms of matching links and of structure, while those falling in the zone III are very different with respect to both characteristics. Networks corresponding to a point in zone II have many common links, but their structure is rather different, while a point in zone IV indicates two networks with few common links, but with similar structure. Full mathematical details about the HIM distance and its two components H and IM are available in \citep{jurman12glocal}. 

\subsection{Stability indicators}
\label{ssec:stability} 
\begin{figure}[!t]
\begin{center}
\fboxrule=1.5mm
\fboxsep=3mm
\fbox{
\begin{minipage}[!h]{0.9\textwidth}
\small
\begin{enumerate}
\item Given a dataset $D$ with $s$ samples and $p$ features, reconstruct (with a chosen algorithm $\textrm{ALG}$) the network $N_D$ on the whole dataset $D$; denote the $p$ nodes of $N_D$ by $x_1^D,\ldots,x_p^D$ and its edges' weight by $a_{hk}^D$, for $k,h=1,\ldots,p$.
\item Choose two integers $n,r$ with $n<s$ and $r\leq \binom{s}{n}$, and build a set $\mathcal{D}_{(n,r)}=\{D_1,\ldots D_r\}$ where $D_i$ is a dataset built choosing $n$ samples from $D$.
\item Reconstruct, by using the same algorithm $\textrm{ALG}$, the corresponding networks $N_{D_i}$ on the subsampled data.
\item Compute the following indicators:
\begin{itemize}
\item $I_1(n,r)=\{\textrm{HIM}(N_D,N_{D_i})\colon i=1,\ldots,r\}$
\item $I_2(n,r)=\{\textrm{HIM}(N_{D_i},N_{D_j})\colon i,j=1,\ldots r, i\not=j\}$
\item $I_3(n,r)=\{a_{hk}^{D_i}\}$ for $i=1,\ldots,r$ and $k,h=1,\ldots,p$
\item $I_4(n,r)=\{\partial(x_{h}^{D_i})\}$ for $i=1,\ldots,r$ and $h=1,\ldots,p$ and $\partial$ the degree function.
\end{itemize}
\item For each set of values $I_i$ compute the mean, the range (defined as the difference between maximum and minumum value) and the 95\% studentized bootstrap confidence intervals \citep{davison97bootstrap} as implemented in the R package \emph{boot} \citep{boot12}.
\item Comparative analysis of the statistics of the four indicators $I_1,\ldots I_4$ will describe the level of confidence (stability) in the network $N_D$, in its links and in its nodes.
\end{enumerate}
\end{minipage}
}
\end{center}
\caption{Definition of the four stability indicators $I_1,\ldots,I_4$.}
\label{fig:defs}
\end{figure}
We introduce now the four stability indicators, for a given subset of the original data and a given number of replicates, producing a set of corresponding inferred networks.
The first two indicators concern the stability of the entire network, measuring the mutual distances of the networks inferred from the different replicates and their distances to the network constructed on the whole dataset.
The other two indicators concern instead the stability (and thus the reliability) of the single nodes and links, in terms of mutual variability of their respective degree and weight.
In Fig.~\ref{fig:defs} we detail the mathematical formulation of the four indicators: the smaller the indicators' values, the stabler the indicators' targets.
In particular, for all experiments on both synthetic and biological datasets we used $n=s-1$, $r=1$ [leave-one-out stability, LOO for short], and $20$ different instances of $k$-fold cross validation (discarding the test portion) for $k=2,4,10$ (denoted by $k2$, $k4$ and $k10$ in what follows), and thus 
$n=\lfloor \frac{s(k-1)}{k} \rfloor$ and $r=20k$.

\section{RESULTS}
\label{sec:results}
\subsection{FDR effect on correlation networks}
\label{ssec:fdr}
As a first experiment, we want to assess the different level of stability in a correlation network inferred by a set of synthetic high-throughput signals when the inference (absolute value of Pearson correlation) is computed with or without False Discovery Rate control (see for instance \citep{jiao11dart}).
As the correlation measure, we use the classical (absolute) Pearson correlation of the WGCNA \citep{horvath11weighted} and the novel correlation measure called Maximal Information Coefficient (MIC), component of the Maximal Information-based Nonparametric Exploration (MINE) statistics \citep{reshef11detecting,speed11correlation,nature12finding}.
For a set of values $n<m$  and an adequate number of resampling $r=\min\{20,\binom{m}{n}\}$, compute the indicators $I_j(n,r)$ for $j=1,\ldots,4$ for all the used algorithms.
\begin{figure}[!t]
\begin{center}
\fboxrule=1.5mm
\fboxsep=3mm
\fbox{
\begin{minipage}[!h]{0.9\textwidth}
\small 
\begin{enumerate}
\item Let $D$ a dataset with $m$ samples described by $q$ features, and let $C(h,k)=|\textrm{cor}(x_h,x_k)|$ where $x_j$ is the $j$-th feature of $D$ across the $m$ samples and $\textrm{cor}$ is a correlation measure.
\item Build the standard correlation network $N_D$ using the rule $a_{hk}=C(h,k)$
\item Build the FDR controlled (at $p$-value $\wp=10^{-z}$) correlation network $M^{\wp}_D$ using the rule
\begin{displaymath}
a_{hk} =
\begin{cases}
C(h,k) & \textrm{if $|F^z_D(h,k)|\leq 1$} \\
0 & \textrm{otherwise, }
\end{cases}
\end{displaymath}
where the set $F_z$ is defined as follows
\begin{equation}
F^z_D = \{ \textrm{cor}(\sigma_i(x_h),\tau_i(x_k))\geq C(h,k) \colon \sigma_i,\tau_i\in S_{m}, i=1,\ldots,\max\{10^z,m!\}\}
\end{equation}
\end{enumerate}
\end{minipage}
}
\end{center}
\caption{Construction of a FDR-corrected correlation network.}
\label{fig:wgcnafdr}
\end{figure}

\subsubsection{Data generation}
\label{sssec:data_synth}
\begin{figure}[!t]
\includegraphics[width=1\textwidth]{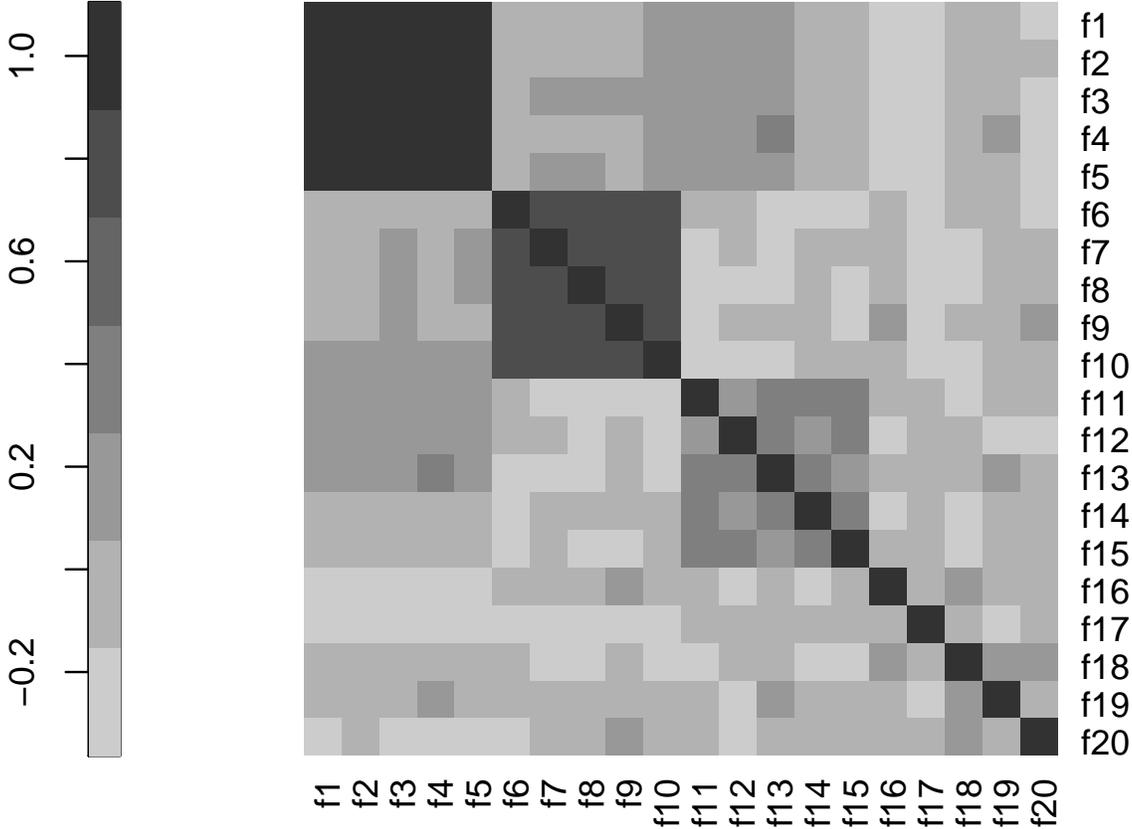}
\caption{The correlation matrix $M_S$ used to generate the synthetic dataset $S$}
\label{fig:MS}
\end{figure}

As a synthetic benchmark for evaluating differences between Pearson and MIC correlation measures, and to assess the impact of the FDR filter on the construction of a correlation network, we built a dataset $S$ consisting of 100 measurements (samples) of 20 variables (features) $f_i$, from which we constructed the corresponding correlation networks on $20$ nodes. 
The dataset $S$ was generated starting from its correlation matrix $M_S$, which was randomly generated with the following three constraints:
\begin{displaymath}
\textrm{Corr}(f_i,f_j)\approx 
\begin{cases}
0.9 & \textrm{for $1\leq i \not= j\leq 5$}\\
0.7 & \textrm{for $6\leq i \not= j\leq 10$}\\
0.4 & \textrm{for $11\leq i \not= j\leq 16$}\ ,
\end{cases}
\end{displaymath} 
for $\textrm{Corr}$ the Pearson correlation.
The correlation matrix $M_S$ is plotted in Fig.~\ref{fig:MS}: clearly, the correlation values in the three groups defined by the above constraints represent true relations between the variables, while all other smaller correlation values are due to the underlying random generation model for $M_S$.

\subsubsection{Results}
\label{sssec:results_synth}
\begin{figure}[!t]
\begin{center}
\begin{tabular}{cc}
\includegraphics[width=.5\textwidth]{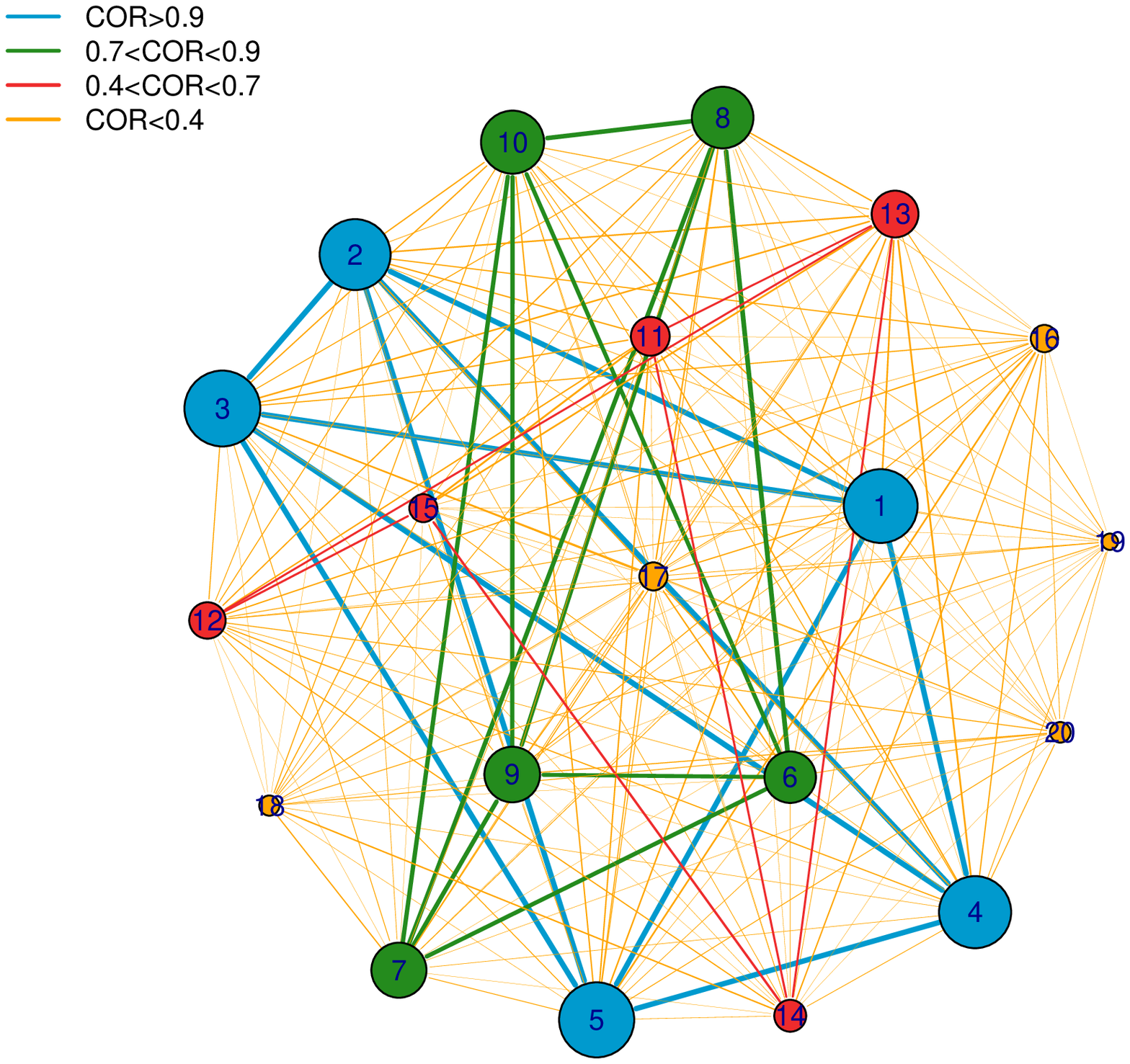} &
\includegraphics[width=.5\textwidth]{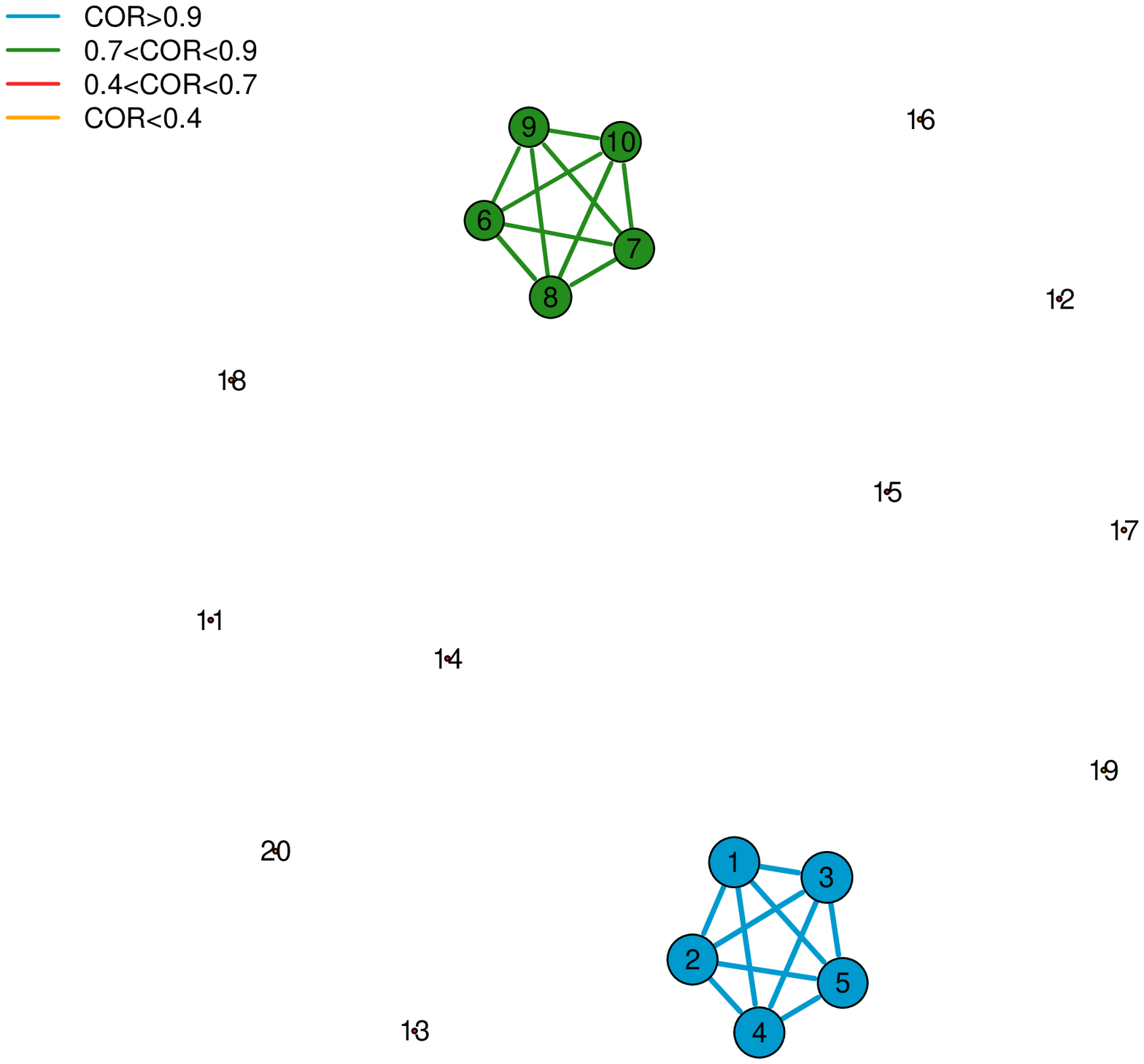} \\
(a) WGCNA & (b) WGCNA FDR 1e-4\\
\multicolumn{2}{c}{\includegraphics[width=.5\textwidth]{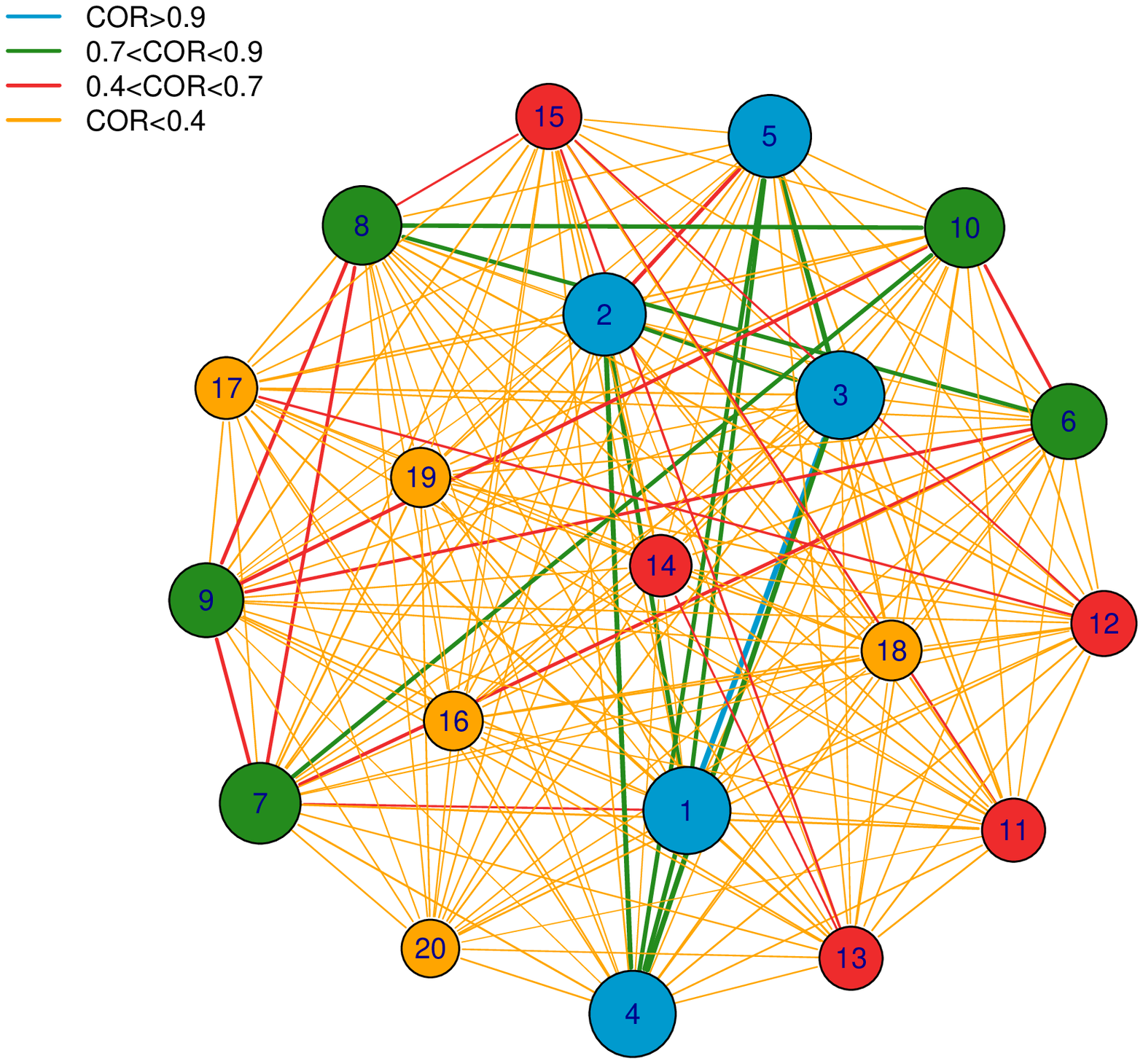}} \\
\multicolumn{2}{c}{(c) MIC}
\end{tabular}
\end{center}
\caption{Correlation networks inferred by the dataset $S$ using (a) absolute Pearson, (b) absolute Pearson with FDR correction at $p$-value $10^{-4}$ and (c) MIC. Node label $i$ corresponds to feature $f_i$, node size is proportional to node degree and link colors identify different classes of link weights.}
\label{fig:w_nets}
\end{figure}

Starting from the dataset $S$ we built five correlation networks, using MIC, absolute Pearson correlation without FDR correction (WGCNA) and absolute Pearson correlation with FDR correction, with $p$-values $\wp=10^{-2},5\cdot 10^{-3}, 10^{-4}$.
The plots of the graphs for three of the networks are displayed in Fig.~\ref{fig:w_nets}.
As expected, while the WGCNA networks with highest FDR correction $\wp=10^{-4}$ is discarding all links as not significant apart from the edges connecting the two disjoint sets of nodes $\{f_i\colon 1\leq i\leq 5\}$ and $\{f_i\colon 6\leq i\leq 11\}$ (the strongest correlations in the matrix $M_S$), WGNCA and MIC generates two fully connected networks with a majority of weak links.
Then we computed the four indicators $I_1, \ldots I_4$ for all the five networks described above, in the setup described in Sec.~\ref{ssec:stability}.
\begin{figure}[!t]
\includegraphics[width=1\textwidth,angle=-90]{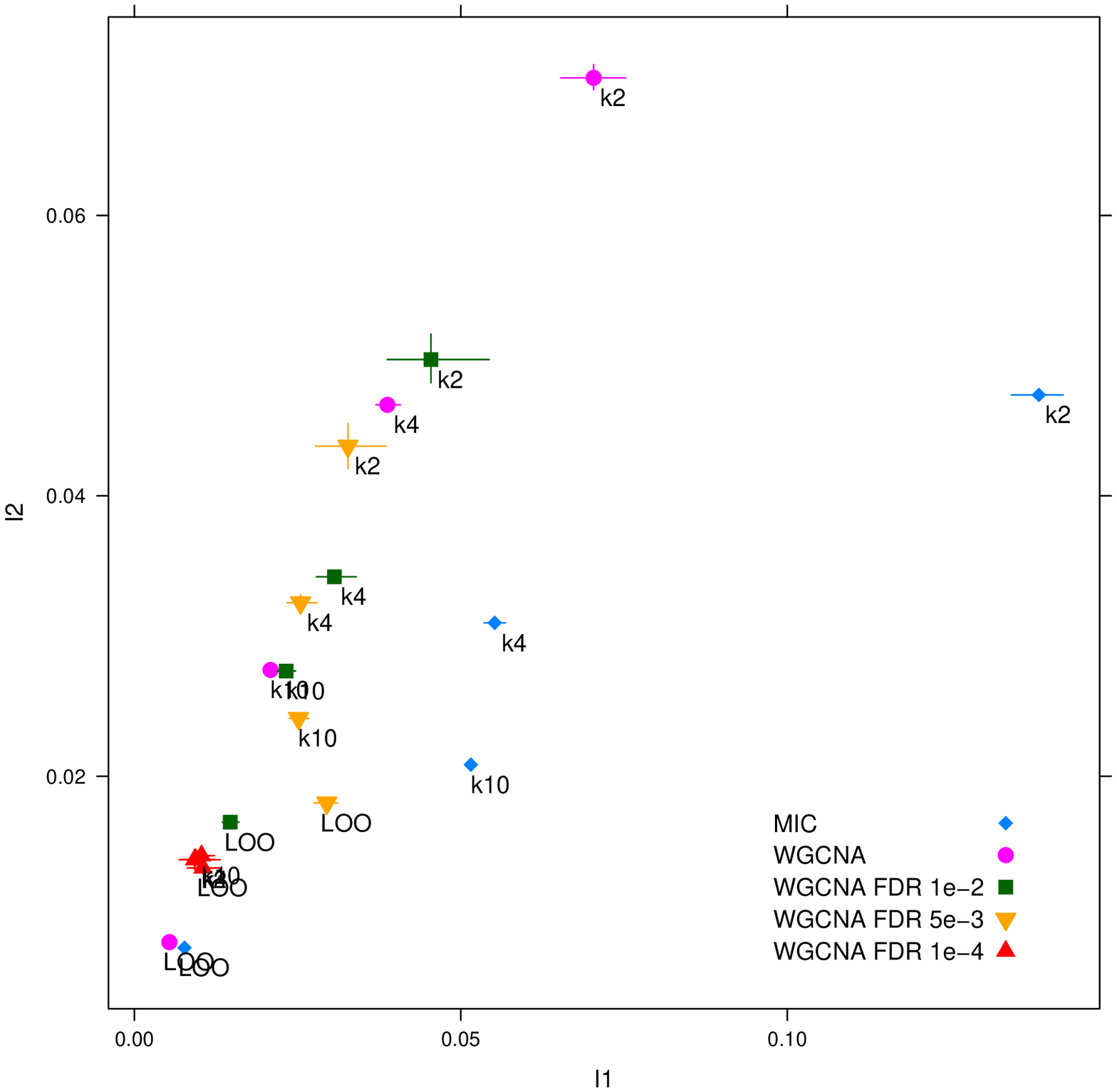}
\caption{$I_1$ and $I_2$ stability indicators (mean and confidence intervals) for different instances of the WGCNA and MIC networks on the dataset $S$ and for different values of data subsampling.}
\label{fig:synth}
\end{figure}
Main statistics for all the indicators $I_1$ and $I_2$ are reported in Tab.~\ref{tab:synth_i1_i2} and displayed in Fig.~\ref{fig:synth}.
\begin{table}[!t]
\caption{Statistics (mean, bootstrap confidence intervals and range) of the stability indicators $I_1$ and $I_2$ for different instances of the WGCNA and MIC networks on the dataset $S$ and for different values of data subsampling.}
\label{tab:synth_i1_i2}
\begin{center}
\scriptsize
\begin{tabular}{lllrrrrr}
\hline
ALG & k & I & mean & CI lower & CI upper & min & max \\ 
\hline
MIC & k10 & $I_1$ & 0.052 & 0.051 & 0.052 & 0.041 & 0.067 \\ 
MIC & k10 & $I_2$ & 0.021 & 0.021 & 0.021 & 0.014 & 0.036 \\ 
MIC & k2 & $I_1$ & 0.139 & 0.134 & 0.142 & 0.112 & 0.158 \\ 
MIC & k2 & $I_2$ & 0.047 & 0.047 & 0.048 & 0.035 & 0.067 \\ 
MIC & k4 & $I_1$ & 0.055 & 0.054 & 0.057 & 0.040 & 0.071 \\ 
MIC & k4 & $I_2$ & 0.031 & 0.031 & 0.031 & 0.022 & 0.045 \\ 
MIC & LOO & $I_1$ & 0.008 & 0.007 & 0.008 & 0.004 & 0.011 \\ 
MIC & LOO & $I_2$ & 0.008 & 0.008 & 0.008 & 0.003 & 0.014 \\ 
WGCNA & k10 & $I_1$ & 0.021 & 0.020 & 0.022 & 0.011 & 0.040 \\ 
WGCNA & k10 & $I_2$ & 0.028 & 0.028 & 0.028 & 0.012 & 0.064 \\ 
WGCNA & k2 & $I_1$ & 0.070 & 0.065 & 0.076 & 0.037 & 0.108 \\ 
WGCNA & k2 & $I_2$ & 0.070 & 0.069 & 0.071 & 0.042 & 0.117 \\ 
WGCNA & k4 & $I_1$ & 0.039 & 0.037 & 0.041 & 0.020 & 0.062 \\ 
WGCNA & k4 & $I_2$ & 0.046 & 0.046 & 0.047 & 0.025 & 0.088 \\ 
WGCNA & LOO & $I_1$ & 0.005 & 0.005 & 0.006 & 0.001 & 0.015 \\ 
WGCNA & LOO & $I_2$ & 0.008 & 0.008 & 0.008 & 0.002 & 0.023 \\ 
WGCNA FDR 1e-2 & k10 & $I_1$ & 0.023 & 0.022 & 0.025 & 0.007 & 0.074 \\ 
WGCNA FDR 1e-2 & k10 & $I_2$ & 0.028 & 0.027 & 0.028 & 0.002 & 0.102 \\ 
WGCNA FDR 1e-2 & k2 & $I_1$ & 0.045 & 0.039 & 0.054 & 0.014 & 0.107 \\ 
WGCNA FDR 1e-2 & k2 & $I_2$ & 0.050 & 0.048 & 0.051 & 0.006 & 0.152 \\ 
WGCNA FDR 1e-2 & k4 & $I_1$ & 0.031 & 0.028 & 0.034 & 0.010 & 0.069 \\ 
WGCNA FDR 1e-2 & k4 & $I_2$ & 0.034 & 0.034 & 0.035 & 0.006 & 0.096 \\ 
WGCNA FDR 1e-2 & LOO & $I_1$ & 0.015 & 0.013 & 0.016 & 0.005 & 0.035 \\ 
WGCNA FDR 1e-2 & LOO & $I_2$ & 0.017 & 0.017 & 0.017 & 0.001 & 0.047 \\ 
WGCNA FDR 5e-3 & k10 & $I_1$ & 0.025 & 0.024 & 0.027 & 0.004 & 0.054 \\ 
WGCNA FDR 5e-3 & k10 & $I_2$ & 0.024 & 0.024 & 0.024 & 0.001 & 0.083 \\ 
WGCNA FDR 5e-3 & k2 & $I_1$ & 0.033 & 0.028 & 0.038 & 0.008 & 0.070 \\ 
WGCNA FDR 5e-3 & k2 & $I_2$ & 0.044 & 0.042 & 0.045 & 0.002 & 0.121 \\ 
WGCNA FDR 5e-3 & k4 & $I_1$ & 0.025 & 0.023 & 0.028 & 0.006 & 0.056 \\ 
WGCNA FDR 5e-3 & k4 & $I_2$ & 0.032 & 0.032 & 0.033 & 0.004 & 0.099 \\ 
WGCNA FDR 5e-3 & LOO & $I_1$ & 0.029 & 0.028 & 0.031 & 0.003 & 0.048 \\ 
WGCNA FDR 5e-3 & LOO & $I_2$ & 0.018 & 0.018 & 0.018 & 0.000 & 0.054 \\ 
WGCNA FDR 1e-4 & k10 & $I_1$ & 0.010 & 0.009 & 0.012 & 0.000 & 0.053 \\ 
WGCNA FDR 1e-4 & k10 & $I_2$ & 0.014 & 0.014 & 0.015 & 0.000 & 0.055 \\ 
WGCNA FDR 1e-4 & k2 & $I_1$ & 0.009 & 0.007 & 0.013 & 0.001 & 0.031 \\ 
WGCNA FDR 1e-4 & k2 & $I_2$ & 0.014 & 0.013 & 0.015 & 0.001 & 0.040 \\ 
WGCNA FDR 1e-4 & k4 & $I_1$ & 0.009 & 0.007 & 0.012 & 0.001 & 0.049 \\ 
WGCNA FDR 1e-4 & k4 & $I_2$ & 0.014 & 0.014 & 0.014 & 0.001 & 0.054 \\ 
WGCNA FDR 1e-4 & LOO & $I_1$ & 0.010 & 0.008 & 0.013 & 0.000 & 0.044 \\ 
WGCNA FDR 1e-4 & LOO & $I_2$ & 0.013 & 0.013 & 0.014 & 0.000 & 0.045 \\ 
\hline
\end{tabular}
\end{center}
\end{table}

As expected, the ratio of the discarded data has a strong impact on both the indicators $I_1$ and $I_2$: in the leave-one-out case the indicators' values are close to zero regardless of the algorithm, while in the $k$-fold cross-validation case the stability is worsening for decreasing values of $k$, in terms of both mean and confidence intervals. This means that the networks inferred from a subset of data have larger distance both mutually and from the network reconstructed from the whole datasets, but also that these distances have larger variability.
From the point of view of the different algorithms involved, the stricter the $p$-value in the FDR controlled WGCNA networks, the stabler the networks, with non controlled WGCNA and MINE as the worst performer in terms of stability. 
This is due to the fact that they are taking into account all possible correlation values, while most of the smaller values do not represent existing relations between variables, but they are rather a noise effect.
As a first result then we showed that the use of a FDR control procedure for correlation help stabilizing the inference procedure, improving the performance of a method already acknowledged as effective \citep{allen12comparing}.

We move now on to discuss the stablest links and nodes in the three cases WGCNA, WGCNA FDR 1e-4 and MIC: in particular, in Tab.~\ref{tab:links} and \ref{tab:nodes} we show the top-ranked links and nodes ordered for decreasing range over mean of their weights across all resampling $k4$.
\begin{table}[!t]
\caption{Top ranked links, ordered by weight range over weight mean across all 20 resampling of $k4$ $4$-fold cross validation, for the three algorithms WGCNA, WGCNAFDR1e-4 and MIC}
\label{tab:links}
\begin{center}
\scriptsize
\begin{tabular}{cr|cr|cr}
\multicolumn{2}{c}{WGCNA} & \multicolumn{2}{|c|}{WGCNA FDR 1e-4} & \multicolumn{2}{c}{MIC} \\
\hline
$f_i-f_j$ & Range/Mean & $f_i-f_j$ & Range/Mean & $f_i-f_j$ & Range/Mean \\
\hline
1 - 3 & 0.03 & 1 - 3 & 0.03 & 	 3 - 4 & 0.20  \\
2 - 3 & 0.04 & 3 - 4 & 0.04 & 	 2 - 3 & 0.20  \\
1 - 2 & 0.04 & 2 - 3 & 0.04 & 	 1 - 3 & 0.21  \\
1 - 4 & 0.04 & 1 - 4 & 0.05 & 	 3 - 5 & 0.22  \\
3 - 4 & 0.04 & 3 - 5 & 0.05 & 	 1 - 2 & 0.23  \\
2 - 4 & 0.04 & 1 - 2 & 0.05 & 	 1 - 5 & 0.25  \\
4 - 5 & 0.04 & 2 - 4 & 0.05 & 	 1 - 4 & 0.26  \\
2 - 5 & 0.05 & 2 - 5 & 0.06 & 	 4 - 5 & 0.27  \\
1 - 5 & 0.05 & 4 - 5 & 0.06 & 	 7 - 10 & 0.28  \\
3 - 5 & 0.05 & 1 - 5 & 0.06 & 	7 - 8 & 0.29  \\
6 - 8 & 0.08 & 6 - 8 & 0.08 & 	 6 - 8 & 0.29  \\
8 - 10 & 0.10 & 7 - 8 & 0.09 & 	   6 - 10 & 0.30  \\
7 - 8 & 0.11 &  8 - 10 & 0.10 & 	   1 - 20 & 0.31  \\
7 - 9 & 0.11 &  8 - 9 & 0.11 & 	   2 - 4 & 0.31  \\
8 - 9 & 0.11 &  6 - 7 & 0.11 & 	   8 - 10 & 0.31  \\
9 - 10 & 0.11 & 7 - 10 & 0.12 & 	   2 - 5 & 0.32  \\
6 - 7 & 0.11 &  7 - 9 & 0.12 & 	   9 - 10 & 0.32  \\
7 - 10 & 0.12 & 9 - 10 & 0.13 & 	   7 - 20 & 0.33  \\
6 - 10 & 0.13 & 6 - 9 & 0.13 & 	   14 - 16 & 0.33  \\
6 - 9 & 0.14 &  6 - 10 & 0.15 & 	   5 - 17 & 0.35  \\
11 - 13 & 0.33 &  & & 	   6 - 7 & 0.35  \\
14 - 15 & 0.41 &  & & 	   11 - 17 & 0.36  \\
13 - 14 & 0.46 & 	   &  & 	   6 - 9 & 0.36  \\
12 - 13 & 0.58 & 	    &  & 	  1 - 10 & 0.37  \\
12 - 15 & 0.60 & 	    &  & 	   10 - 11 & 0.37  \\
11 - 14 & 0.62 & 	    &  & 	   10 - 20 & 0.37  \\
13 - 15 & 0.71 & 	    &  & 	   4 - 17 & 0.37  \\
11 - 15 & 0.78 & 	    &  & 	   2 - 8 & 0.37  \\
14 - 18 & 0.78 & 	    &  & 	   4 - 10 & 0.37  \\
3 - 11 & 0.83 & 	    &  & 	   6 - 13 & 0.37  \\
5 - 11 & 0.83 & 	    &  & 	   2 - 14 & 0.37  \\
1 - 11 & 0.84 & 	    &  & 	   9 - 11 & 0.38  \\
4 - 11 & 0.85 & 	     &  & 	   15 - 16 & 0.38  \\
3 - 10 & 0.87 & 	     &  & 	   15 - 17 & 0.38  \\
5 - 16 & 0.89 & 	     &  & 	   7 - 13 & 0.39 \\
8 - 17 & 0.89 & 	     &  & 	   9 - 18 & 0.39  \\
2 - 11 & 0.91 & 	     &  & 	   12 - 19 & 0.39  \\
8 - 12 & 0.91 & 	     &  & 	   6 - 18 & 0.39  \\
4 - 13 & 0.91 & 	     &  & 	   8 - 9 & 0.39  \\
1 - 13 & 0.93 & 	     &  & 	   4 - 18 & 0.39  \\
3 - 13 & 0.93 & 	     &  & 	   16 - 17 & 0.39  \\
8 - 13 & 0.94 & 	     &  & 	   4 - 19 & 0.39  \\
9 - 17 & 0.94 & 	     &  & 	   16 - 19 & 0.39 \\
1 - 16 & 0.95 & 	     &  & 	   7 - 19 & 0.40  \\
1 - 10 & 0.95 & 	     &  & 	   5 - 8 & 0.40  \\
14 - 16 & 0.97 & 	     &  & 	   14 - 15 & 0.40  \\
5 - 10 & 0.97 & 	     &  & 	   13 - 15 & 0.40  \\
11 - 12 & 0.98 & 	     &  & 	   4 - 11 & 0.40  \\
12 - 16 & 0.98 & 	     &  & 	   7 - 9 & 0.41  \\
2 - 13 & 0.99 & 	     &  & 	   13 - 19 & 0.41 \\
\hline
\end{tabular}
\end{center}
\end{table}
The results collected in the tables are consistent with the structure of the starting correlation matrix $M_S$ and the behaviour of the inference algorithms.
For the WGCNA case, the top $20$ stablest links are those of the two fully connected subgroups $F_{1,5}=\{f_i\colon 1\leq i\leq 5\}$ and $F_{6,10}\{f_i\colon 6\leq i\leq 10\}$ with largest Pearson correlation values in $M_S$. 
The same applies to WGCNA FDR 1e-4 (and with approximately the same values of weight range over weight mean as for WGCNA), for which these 20 links are the only existing (see Fig.~\ref{fig:w_nets}).
Among the following ranked links in WGCNA, those belonging to the $F_{11,15}=\{f_i\colon 11\leq i\leq 15\}$ group (whose correlation of about 0.3 was imposed as a constraint for $M_S$) are emerging, with a couple of exceptions, but with larger instability values (0.33-0.78 vs. 0.03-0.14).
The remaining links are the unstablest, displaying Range/Mean values always larger than 0.83: they are the randomly correlated links of $M_S$.
It is interesting to note that the MIC network, due to the nature of the MIC statistics aimed at detecting relations between variables other than linear, displays similar but not identical results: the values of Range/Mean are confined in a narrower interval, and, although many links belonging to the $F_{1,5}$ and $F_{6,10}$ groups are highly ranked, some of them can also be found in much lower positions of the standing.

Similar considerations hold for the ranking of the stablest nodes: for WGCNA, the top ranking nodes are the $F_{1,5}$ and the $F_{6,10}$ (with similar Range/Mean values), with those in $F_{11,15}$ come next, leaving the remaining five as the most unstable, with higher Range/Mean values.
These five nodes, on the contrary, are the stablest for WGCNA FDR 1e-4: in fact, they are not wired to any other node in any of the resampling, so their Range/Mean values are void.
The nodes $F_{1,5}\cup F_{6,10}$ then follow in the ranking with small associated values, and the nodes $F_{11,15}$ close the standing with definitely higher values. 
In fact, although the nodes $F_{11,15}$ have degree zero in the WGCNA FDR 1e-4 inferred from the whole $S$, some links involving them exist in some of the resampling on the subset of data.
To conclude with, in the MIC case again the ranking values span a much narrower range than the other two cases, and the obtained dwranking has most of the nodes in $F_{1,5}$ in top positions, while for the other nodes the relation with the structure of $M_S$ is very weak.

Finally, the analogous tables for other ratios of the data subsampling schema (LOO, $k2$ and $k10$) are almost identical.
\begin{table}[!t]
\caption{Top ranked nodes, ordered by degree range over degree mean across all 20 resampling of $k4$ $4$-fold cross validation, for the three algorithms WGCNA, WGCNA FDR 1e-4 and MIC. (*) indicates that Ratio and Mean are both zero.}
\label{tab:nodes}
\begin{center}
\scriptsize
\begin{tabular}{cr|cr|cr}
\multicolumn{2}{c}{WGCNA} & \multicolumn{2}{|c|}{WGCNA FDR 1e-4} & \multicolumn{2}{c}{MIC} \\
\hline
$f_i$ & Range/Mean & $f_i$ & Range/Mean & $f_i$ & Range/Mean \\
\hline
4 &  0.17 & 	   16 & 0* & 	 3 & 0.08 \\
10 &  0.18 & 	   17 & 0* & 	 19 & 0.08 \\
3 & 0.20 & 	   18 & 0* & 	 1 & 0.08 \\
1 & 0.21 & 	   19 & 0* & 	 4 & 0.09 \\
9 & 0.23 & 	   20 & 0* & 	 8 & 0.09 \\
2 & 0.23 & 	    3 & 0.03 &  10 & 0.09 \\
5 & 0.24 & 	  1 & 0.04 & 	  5 & 0.10 \\
7 & 0.24 & 	  2 & 0.04 & 	  2 & 0.10 \\
6 & 0.24 & 	  5 & 0.05 & 	  17& 0.10 \\
8 & 0.25 & 	  7 & 0.07 & 	  20& 0.10 \\
11 & 0.40 & 	  8 & 0.07 & 	  15& 0.11 \\
13 & 0.40 & 	  6 & 0.09 & 	  9 & 0.11 \\
15 & 0.43 & 	  9 & 0.09 & 	  13& 0.11 \\
12 & 0.45 & 	   10 & 0.09 &   11& 0.11 \\
14 & 0.48 & 	  4 & 0.13 & 	  16& 0.11 \\
18 & 0.55 & 	   15 & 4.42 &   12& 0.11 \\
16 & 0.60 & 	   14 & 7.05 &   7&  0.11 \\
17 & 0.68 & 	   12 & 22.82 &  6 & 0.12\\
20 & 0.70 & 	   13 & 26.05 &  14& 0.13 \\
19 & 1.15 & 	   11 & 41.83 &  18& 0.13 \\
\hline
\end{tabular}
\end{center}
\end{table}

\subsection{miRNA network on a Hepatocellular Carcinoma dataset}
\label{ssec:bio}
Investigating the relations connecting human microRNA (miRNA) and how they evolve in cancer has been recently a key topic for researcher in biology \citep{volinia10reprogramming,bandyopadhyay10development}, with hepatocellular carcinoma (HCC) as a notable example \citep{law11emerging,gu12gene}.
In the following example, we use the stability indicators $I_1,\ldots,I_4$ on a recent miRNA microarray dataset with two phenotypes to highlight differences in the corresponding inferred networks.
As reconstruction algorithm we use the Context Likelihood of Relatedness (CLR) approach \citep{faith07large}, belonging to the relevance networks class of algorithms and generating undirected weighted graphs with weights bounded between zero and one.
In particular, interactions are scored by using the mutual information between the corresponding gene expression levels coupled with an adaptive background correction step.
Although suboptimal if the number of variables is much larger than the number of variables, it was observed that CLR performes well in terms of prediction accuracy and some CLR predictions in literature were later experimentally validated \citep{ambroise12transcriptional}.

\subsubsection{Data description}
\label{sssec:data_hcc}
We start out from the Hepatocellular Carcinoma dataset introduced in the paper \citep{budhu08identification} and later used in \citep{ji09microrna}, publicly available at the Gene Expression Omnibus (GEO, \url{http://www.ncbi.nlm.nih.gov/geo/}) at the accession number GSE6857.
The dataset collects 482 tissue samples from 241 patients affected by hepatocellular carcinoma (HCC). 
For each patients, a sample from cancerous hepatic tissue and a sample from surrounding non-cancerous hepatic tissue are available, hybridized on the Ohio State University CCC MicroRNA Microarray Version 2.0 platform consisting of 11520 probes collecting expressions of 250 non-redundant human and 200 mouse microRNA (miRNA).
After a preprocessing phase including imputation of missing values as in \citep{troyanskaya01missing} and discarding probes corresponding to non-human (mouse and controls) miRNA, we end up with the dataset $\mathcal{HCC}$ of 240+240 paired samples described by 210 human miRNA,
with the cohort consisting of 210 male and 30 female patients.
We thus parted the whole dataset $\mathcal{HCC}$ into four subsets combining the sex and disease status phenotypes, collecting respectively the cancer tissue for the male patients (MT), the cancer tissue for the female patients (FT) and the corresponding two datasets including the non cancer tissues (MnT, FnT).

\subsubsection{Results}
\label{sssec:results_hcc}
\begin{figure}[!h]
\begin{center}
\begin{tabular}{lcr}
\raisebox{3cm}{
\begin{tabular}{ccc|c}
MnT & FT & FnT \\
\hline
0.0412  & 0.0858  & 0.0235 & MT \\
        & 0.1265  & 0.0618 & MnT \\
	& 	  & 0.0684 & FT \\
\end{tabular}
} & \phantom{aaaa} & 
\includegraphics[width=0.45\textwidth]{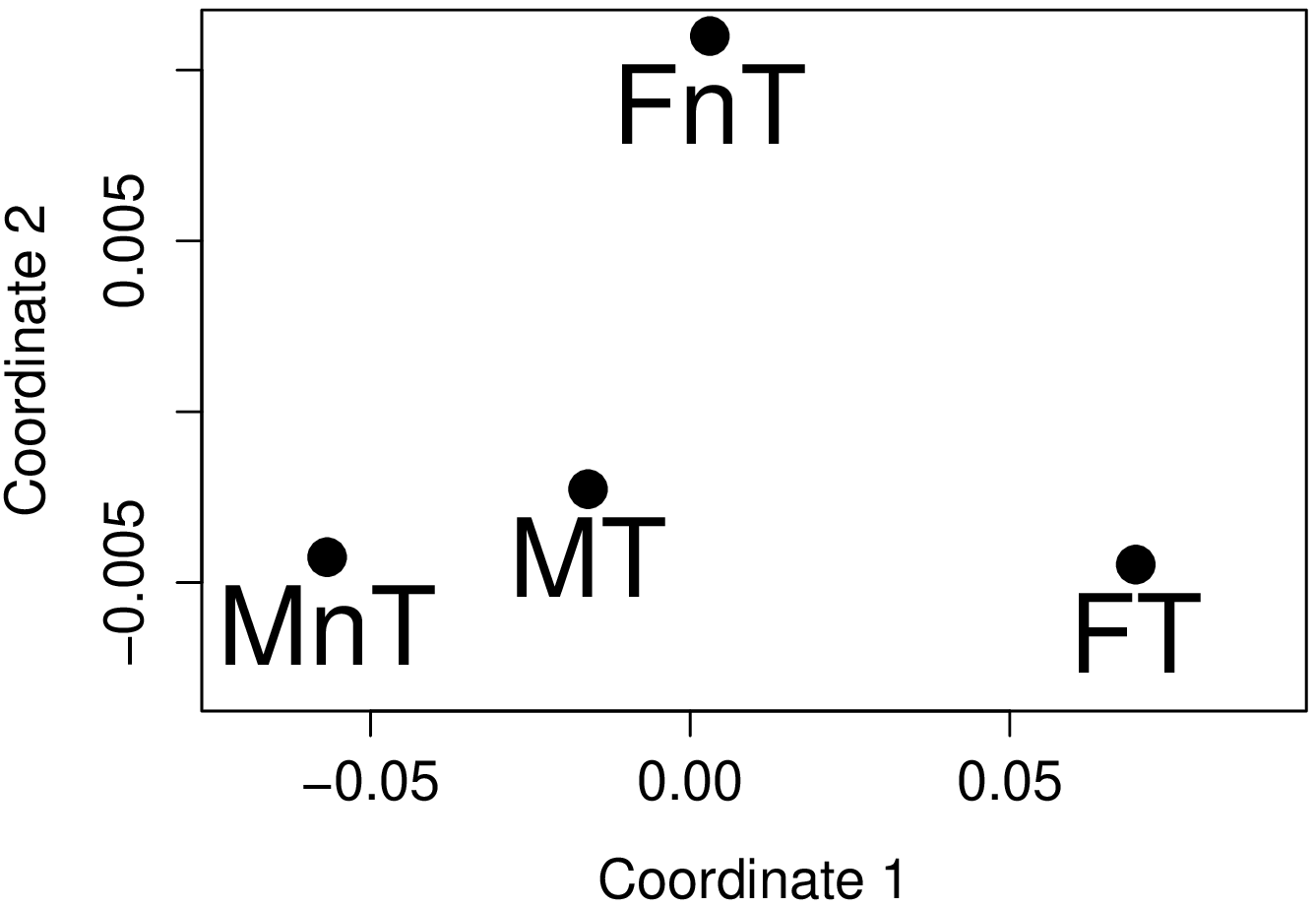}
\end{tabular}
\end{center}
\caption{Mutual HIM distances for the four CLR inferred networks MT, MnT, FT, FnT reconstructed from the whole corresponding subsets and corresponding 2D multidimensional scaling plot.}
\label{fig:hcc_him}
\end{figure}

\begin{figure}[!t]
\begin{center}
\begin{tabular}{cc}
\includegraphics[width=0.5\textwidth]{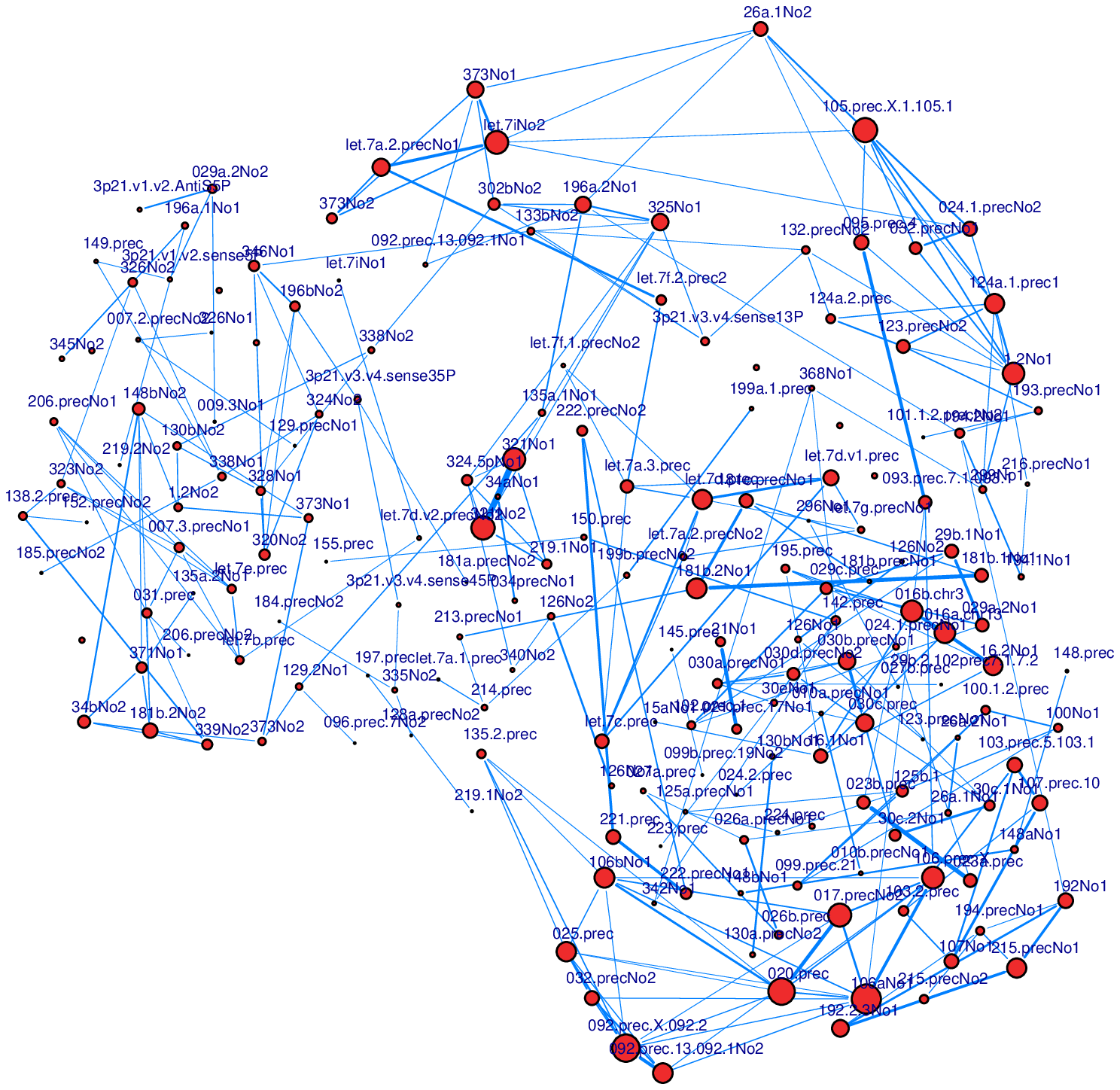} &
\includegraphics[width=0.5\textwidth]{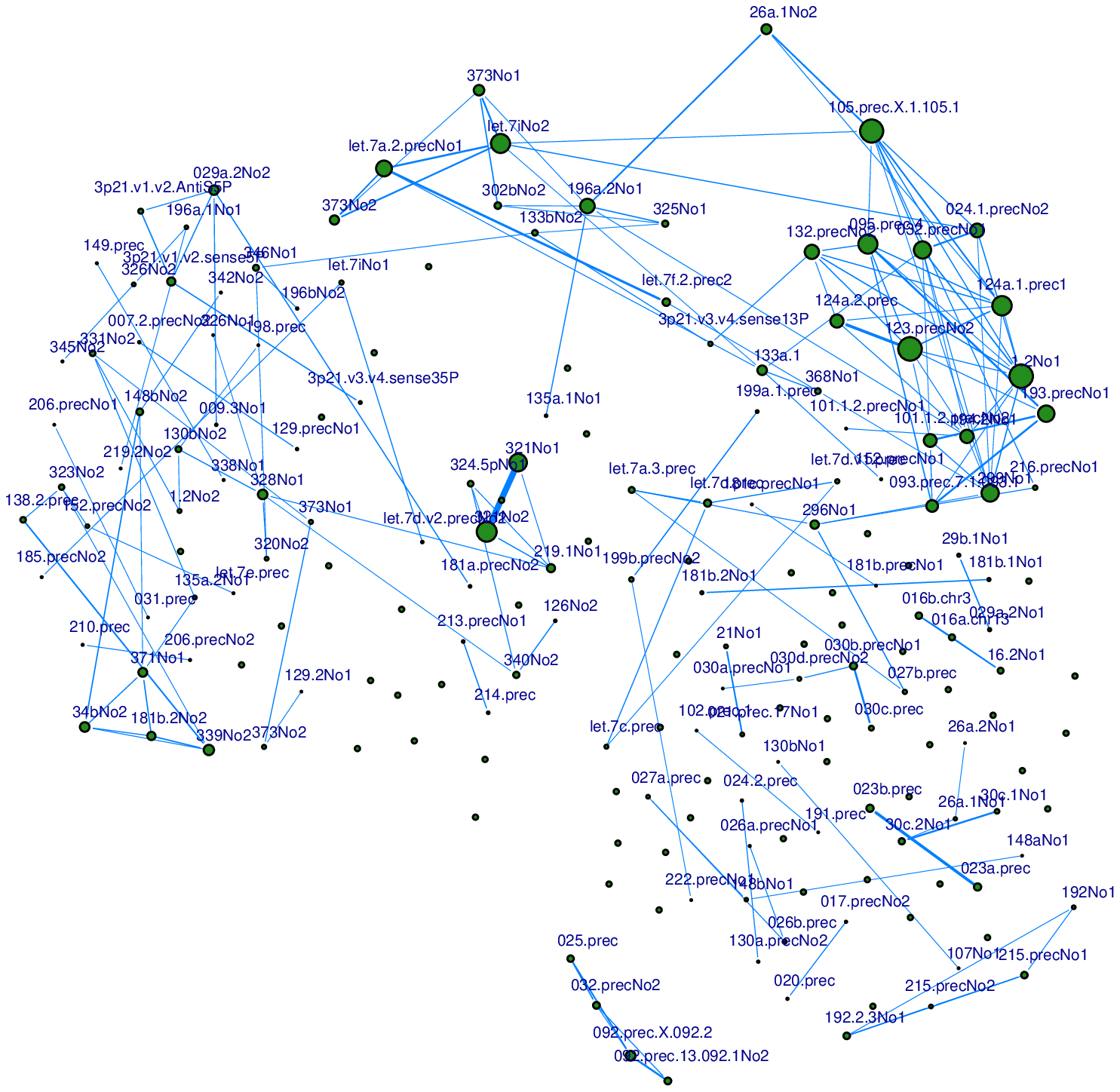} \\
(a) MT, $d=0.0153$ & (b) MnT, $d=0.0092$ \\
\includegraphics[width=0.5\textwidth]{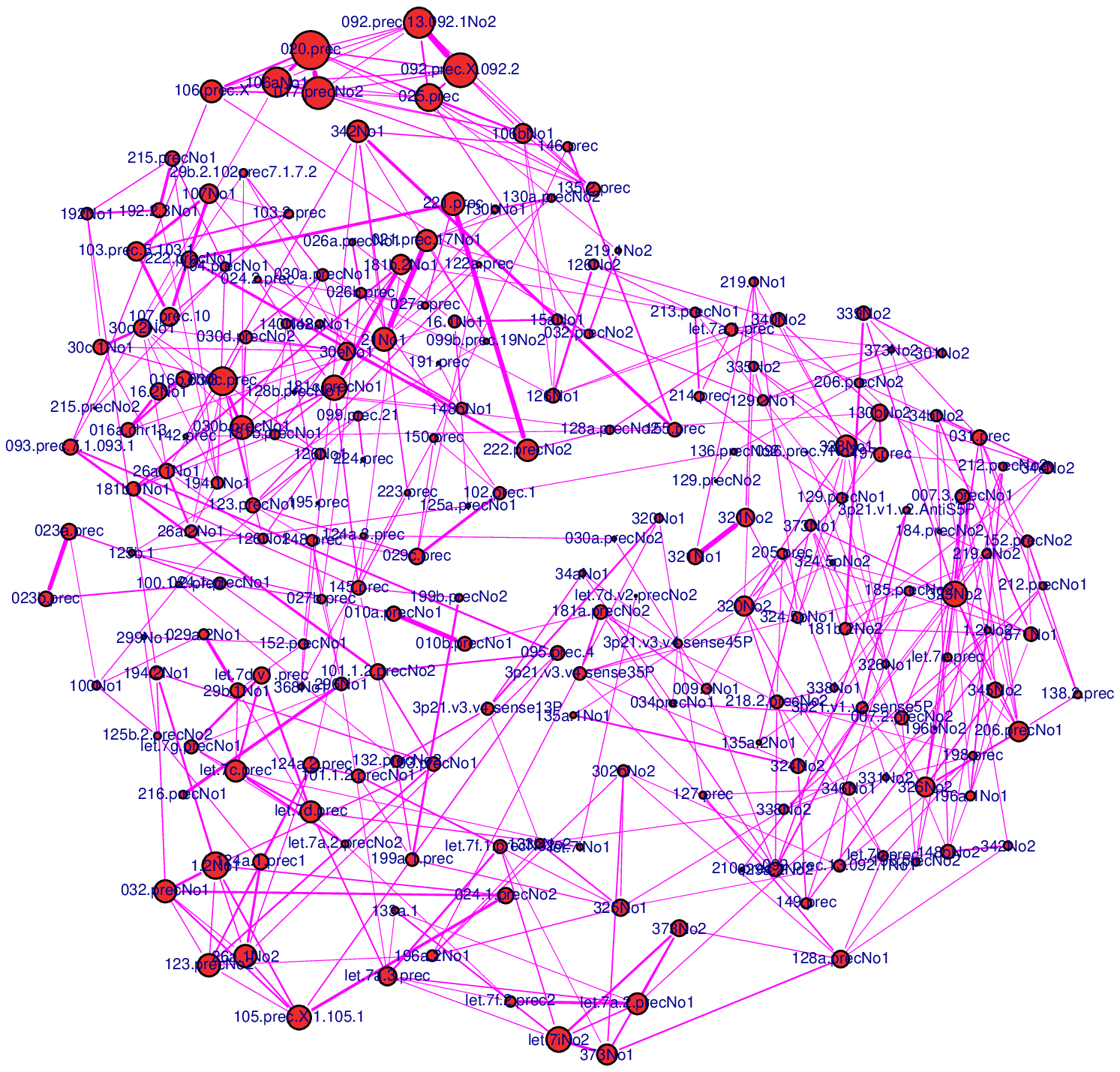} &
\includegraphics[width=0.5\textwidth]{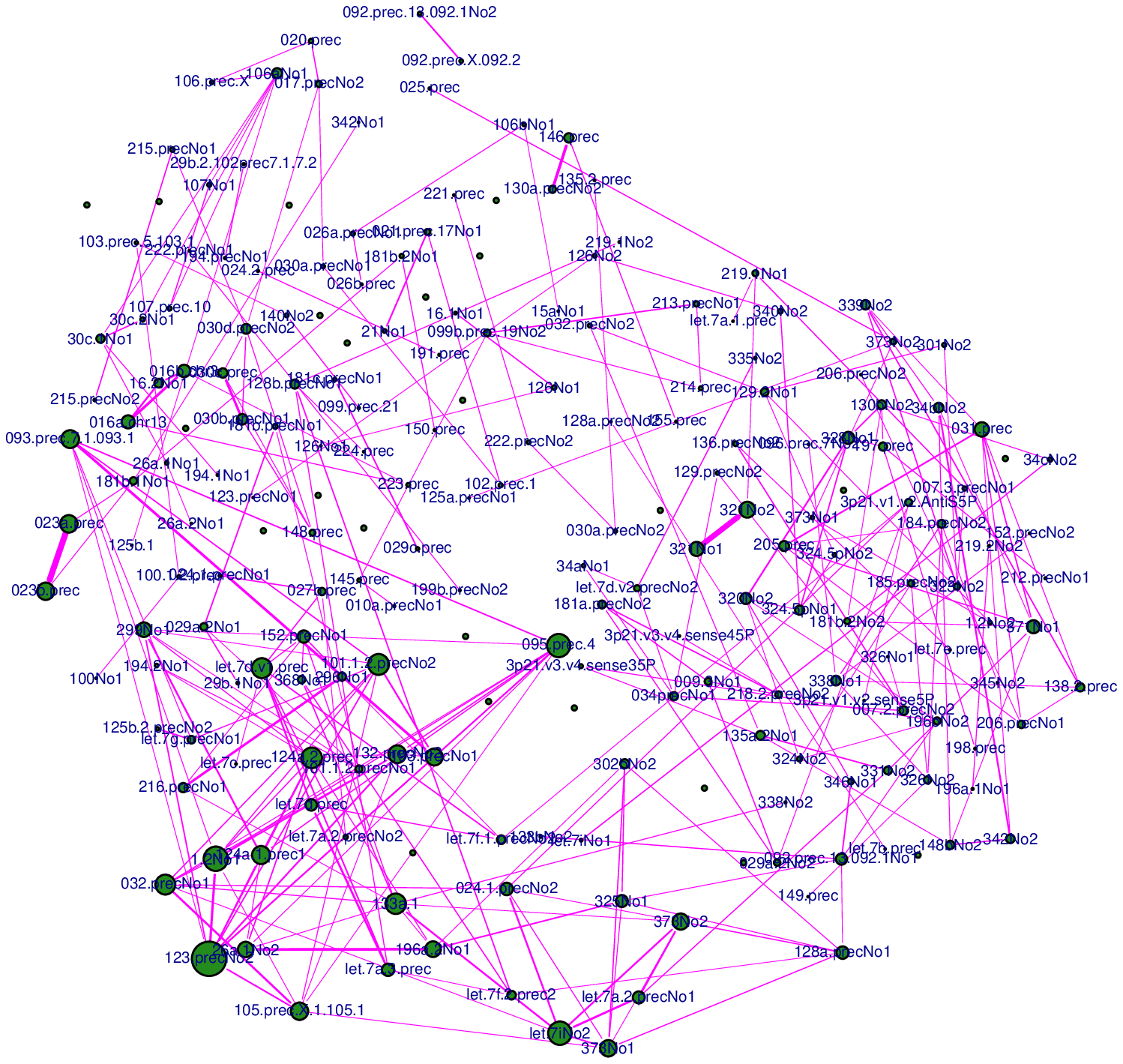} \\
(c) FT, $d=0.0206$ & (d) FnT, $d=0.0121$  \\
\end{tabular}
\caption{CLR networks (and corresponding density values) inferred from the 4 subsets (a) Male Tumoral (MT) (b) Male not Tumoral (MnT) (c) Female Tumoral (FT) and (d) Female non Tumoral (FnT) of the datasets $\mathcal{HCC}$. Links are thresholded at weight 0.1, node position is fixed across the four networks, node dimension is proportional to the degree and edge width is proportional to link weight.}
\label{fig:hcc_nets}
\end{center}
\end{figure}

\begin{figure}[!t]
\includegraphics[width=1\textwidth,angle=-90]{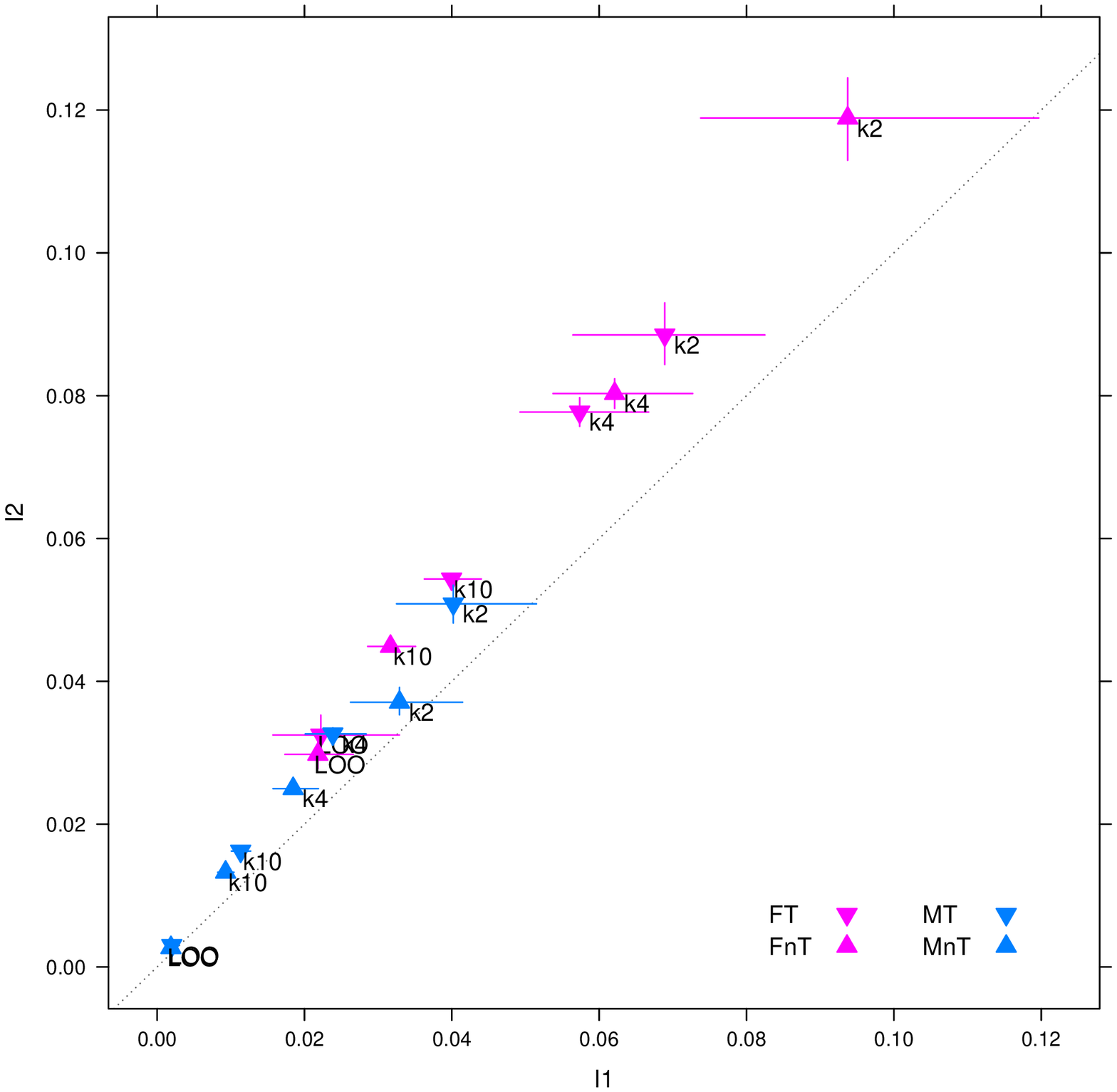}
\caption{$I_1$ and $I_2$ stability indicators (mean and confidence intervals) of CLR inferred networks for different values of data subsampling on the four subgroups Male Tumoral (MT), Male not Tumoral (MnT), Female Tumoral (FT) and Female non Tumoral (FnT) of the datasets $\mathcal{HCC}$.}
\label{fig:hcc}
\end{figure}

Using the CLR algorithm we first generated the four networks inferred from the whole sets of data and corresponding to the combinations of the two binary phenotypes: a portrait of the resulting graphs is depicted in Fig.~\ref{fig:hcc_nets}, discarding links whose weight is smaller than 0.1.
As a first observation, the four networks have a different structure, for instance the tumoral tissues graphs being more connected than the controls and the female graphs more than the corresponding male ones (see for instance the density values in Fig.~\ref{fig:hcc_nets}).
In particular, their mutual HIM distances are reported in Tab.~\ref{fig:hcc_him}, together with the corresponding two-dimensional scaling plot, showing that the networks corresponding to the female patients (and, in particular, the one inferred from cancer tissue) are notably different from those arising from the subset of data for the male patients.
We then computed the stability indicators $I_1$ and $I_2$ in the setup described in Sec.~\ref{ssec:stability}, and the corresponding statistics are collected and displayed in Tab.~\ref{tab:hcc_i1_i2} and Fig.~\ref{fig:hcc}. 

It is immediately evident the different sample size impact on the network stability: the networks corresponding to male patients have smaller values for $I_1$ and $I_2$ (and thus they are much stabler) than the corresponding female counterparts, and this effect is even stronger than the one due to the ratio of the chosen subsets of data: the leave-one-out stability for FT and FnT is worse than k10 and k4 stability for MT and MnT. 
On the other hand, while control and cancer networks display similar level of stability in the male networks at all levels of subsampling ratio, in the female group the network associated to the controls is much stabler than the matching control networks, and this is evident when the size of the subset used for inference gets smaller, in particular for $k=2$.

\begin{table}[!t]
\caption{Statistics (mean, bootstrap confidence intervals and range) of the stability indicators $I_1$ and $I_2$ for the CLR inferred networks on the datasets MT, MnT, FT, FnT, for different values of data subsampling.}
\label{tab:hcc_i1_i2}
\begin{center}
\scriptsize
\begin{tabular}{lllrrrrr}
\hline
PROBL & k & I & mean & lower & upper & min & max \\ 
\hline
FT & k10 & $I_1$ & 0.040 & 0.037 & 0.044 & 0.002 & 0.177 \\ 
FT & k10 & $I_2$ & 0.054 & 0.054 & 0.055 & 0.000 & 0.256 \\ 
FT & k2 & $I_1$ & 0.069 & 0.056 & 0.082 & 0.006 & 0.154 \\ 
FT & k2 & $I_2$ & 0.089 & 0.084 & 0.093 & 0.005 & 0.250 \\ 
FT & k4 & $I_1$ & 0.057 & 0.049 & 0.066 & 0.004 & 0.190 \\ 
FT & k4 & $I_2$ & 0.078 & 0.076 & 0.080 & 0.003 & 0.305 \\ 
FT & LOO & $I_1$ & 0.022 & 0.016 & 0.032 & 0.002 & 0.093 \\ 
FT & LOO & $I_2$ & 0.032 & 0.030 & 0.035 & 0.001 & 0.143 \\ 
FnT & k10 & $I_1$ & 0.032 & 0.029 & 0.035 & 0.002 & 0.093 \\ 
FnT & k10 & $I_2$ & 0.045 & 0.044 & 0.045 & 0.000 & 0.179 \\ 
FnT & k2 & $I_1$ & 0.094 & 0.071 & 0.117 & 0.006 & 0.257 \\ 
FnT & k2 & $I_2$ & 0.119 & 0.113 & 0.124 & 0.006 & 0.391 \\ 
FnT & k4 & $I_1$ & 0.062 & 0.054 & 0.072 & 0.005 & 0.203 \\ 
FnT & k4 & $I_2$ & 0.080 & 0.078 & 0.082 & 0.003 & 0.307 \\ 
FnT & LOO & $I_1$ & 0.022 & 0.017 & 0.027 & 0.003 & 0.048 \\ 
FnT & LOO & $I_2$ & 0.030 & 0.028 & 0.032 & 0.001 & 0.094 \\ 
MT & k10 & $I_1$ & 0.011 & 0.010 & 0.013 & 0.001 & 0.048 \\ 
MT & k10 & $I_2$ & 0.016 & 0.016 & 0.016 & 0.001 & 0.092 \\ 
MT & k2 & $I_1$ & 0.040 & 0.033 & 0.051 & 0.003 & 0.146 \\ 
MT & k2 & $I_2$ & 0.051 & 0.048 & 0.054 & 0.003 & 0.218 \\ 
MT & k4 & $I_1$ & 0.024 & 0.020 & 0.029 & 0.002 & 0.099 \\ 
MT & k4 & $I_2$ & 0.033 & 0.032 & 0.033 & 0.001 & 0.148 \\ 
MT & LOO & $I_1$ & 0.002 & 0.002 & 0.002 & 0.000 & 0.018 \\ 
MT & LOO & $I_2$ & 0.003 & 0.003 & 0.003 & 0.000 & 0.030 \\ 
MnT & k10 & $I_1$ & 0.009 & 0.008 & 0.010 & 0.001 & 0.034 \\ 
MnT & k10 & $I_2$ & 0.013 & 0.013 & 0.013 & 0.001 & 0.061 \\ 
MnT & k2 & $I_1$ & 0.033 & 0.026 & 0.041 & 0.003 & 0.104 \\ 
MnT & k2 & $I_2$ & 0.037 & 0.035 & 0.039 & 0.002 & 0.158 \\ 
MnT & k4 & $I_1$ & 0.018 & 0.015 & 0.022 & 0.001 & 0.067 \\ 
MnT & k4 & $I_2$ & 0.025 & 0.024 & 0.026 & 0.001 & 0.102 \\ 
MnT & LOO & $I_1$ & 0.002 & 0.002 & 0.002 & 0.000 & 0.009 \\ 
MnT & LOO & $I_2$ & 0.003 & 0.003 & 0.003 & 0.000 & 0.016 \\ 
\hline
\end{tabular}
\end{center}
\end{table}

Finally, to show how to use indicators $I_3$ and $I_4$ to extract information about stability of some interesting links, we first rank all links according to their weight Range/Mean value for all the four cases MT, MnT, FT, FnT, and then we point out six  links worth a comment, listed in Tab.~\ref{tab:mirna}.
The link (a) is top ranking in all four cases as expected, since \emph{hsa-mir\_321No1} and \emph{hsa-mir\_321No2} denote essentially the same miRNA (identical or with very similar sequences, \citep{ambros03uniform}.
The same applies to the links (b) and (c), but in these cases the stability of these two links in the FnT network is not as good as in the other three cases, probably due to the presence of noise in the data.
The link (d) is interesting because of the difference of its stability between the male and the female networks, indicating a link probably associated to sex rather than HCC.
The behaviour of link (e) is even more singular: it is one of the stablest links for the FT network, while is not even picked up as a link by CLR in the FnT network.
Finally, link (f) is a very well known connection in literature, strongly associated to cancer \citep{volinia10reprogramming,braun08p53,georges08coordinated} as confirmed by its high stability in the MT and FT networks only.
 
\begin{table}[!t]
\caption{Position in the weight Range/Mean ranking in the four cases MT, MnT, FT, FnT for six miRNA-miRNA links.}
\label{tab:mirna}
\begin{center}
\begin{tabular}{crrrrrrll}
  \hline
id & hsa-mir\_idx1 & hsa-mir\_idx2 & MT & MnT & FT & FnT \\		
\hline
(a) & 321No1 & 321No2 & 1 & 1 & 9 & 2	\\	
(b) & 016b.chr3 & 16.2No1  3 & 12 & 15 & 309 \\ 	 
(c) &021.prec.17No1 & 21No1 & 27 & 5 & 2 & 921 \\ 
(d) &219.1No1 & 321No2 & 2 & 6 & 1903 & 314 		\\  
(e) &326No1 & 342No2  & 132 & 1017 & 3 & - 		\\ 
(f) &192.2.3No1 & 215.precNo1 & 4 & 300 & 4 & 3340 	\\ 
   \hline
\end{tabular}
\end{center}
\end{table}

\section{CONCLUSIONS}
\label{sec:conclusions}
We introduced a suite of four stability indicators for assessing the variability of network reconstruction algorithm as functions of a data subsampling procedure.
The aim here is to provide the researchers with an effective tool to compare either the inference algorithms or the investigated dataset.
Two indicators are based on a measure of a normalized distance between networks and they are global, giving a confidence measure on the whole inferred dataset, while the other two are local, associating a reliability score to the network nodes and detected links.
They are of particular interest when no gold standard is known for the studied task, so they can work as a substitute for the algorithm accuracy.
We demonstrated their consistency on a synthetic dataset, and we showed their use on a high-throughput microarray experiment, with two widely known inference methods such as WGCNA and CLR.

\section*{ACKNOWLEDGEMENTS}
The authors acknowledge funding by the European Union FP7 Project HiperDART.

\section*{DISCLOSURE STATEMENT}
No competing financial interests exist.

\bibliographystyle{jcbnat}    
\bibliography{jurman12stability}
\begin{flushright}
\begin{tabular}{r}
Address correspondence to:\\
\\
\textit{Giuseppe Jurman}\\
\textit{Fondazione Bruno Kessler (FBK)}\\
\textit{via Sommarive 18 - Povo}\\
\textit{I-38123 Trento}\\
\textit{Italy}\\
\\
\textit{E-mail:} jurman@fbk.eu
\end{tabular}
\end{flushright}
\end{document}